\documentclass{aa}  

\usepackage{graphicx}
\usepackage{natbib}
\usepackage{epstopdf}  
\usepackage{lscape}
\usepackage{booktabs}
\usepackage{txfonts}
\usepackage{amsmath}
\usepackage{algorithmic}
\usepackage{bm}
\usepackage{mathrsfs}
\usepackage{color}
\usepackage{subfigure}
\usepackage{float}

\usepackage[colorlinks,linkcolor=blue,anchorcolor=blue,citecolor=blue]{hyperref}

\begin{document} 

\title{Studying the properties of reconnection-driven turbulence}

\author{Shi-Min Liang \inst{1,2}
        \and
        Jian-Fu Zhang \inst{2,3}
        \and
        Na-Na Gao \inst{2}
        \and
        Nian-Yu Yi \inst{1,4,5}
        }
\institute{School of Mathematics and Computational Sciences, Xiangtan University, Xiangtan, Hunan 411105, People’s Republic of China\\
\and Department of Physics, Xiangtan University, Xiangtan, Hunan 411105, People’s Republic of China\\
\email{jfzhang@xtu.edu.cn (JFZ)}\\
\and Key Laboratory of Stars and Interstellar Medium, Xiangtan University, Xiangtan 411105, People’s Republic of China\\
\and Hunan Key Laboratory for Computation and Simulation in Science and Engineering, Xiangtan 411105, People’s Republic of China\\
\and Hunan National Center for Applied Mathematics, Xiangtan 411105, People’s Republic of China\\
\email{yinianyu@xtu.edu.cn (NYY)}\\
}
   \date{Received XXX; accepted XXX}

 \abstract
  {
  Magnetic reconnection, often accompanied by turbulence interaction, is a ubiquitous phenomenon in astrophysical environments. However, the current understanding of the nature of turbulent magnetic reconnection remains insufficient.
  }
  {
  We investigate the statistical properties of reconnection turbulence in the framework of the self-driven reconnection.
  }
  {
Using the open-source software package AMUN, we first perform numerical simulations of turbulent magnetic reconnection. We then obtain the statistical results of reconnection turbulence by traditional statistical methods such as the power spectrum and structure function.
  }
 {
  Our numerical results demonstrate: (1) the velocity spectrum of reconnection turbulence follows the classical Kolmogorov type of $E\propto k^{-5/3}$, while the magnetic field spectrum is steeper than the Kolmogorov spectrum, which are independent of limited resistivity, guide field, and isothermal or adiabatic fluid states; (2) most of the simulations show the anisotropy cascade, except that the presence of a guide field leads to an isotropic cascade; (3) reconnection turbulence is incompressible in the adiabatic state, with energy distribution dominated by the velocity solenoidal component; (4) different from pure magnetohydrodynamic (MHD) turbulence, the intermittency of the velocity field is stronger than that of the magnetic field in reconnection turbulence.
  }
 { 
The steep magnetic field spectrum, together with the velocity spectrum of Kolmogorov type, can characterize the feature of the reconnection turbulence. In the case of the presence of the guide field, the isotropy of the reconnection turbulence cascade is also different from the cascade mode of pure MHD turbulence. Our experimental results provide new insights into the properties of reconnection turbulence, which will contribute to advancing the self-driven reconnection theory. 
}

\maketitle

\section{Introduction} \label{sec:intro}
Magnetic reconnection is a fundamental physical process that facilitates the conversion of magnetic energy into kinetic and internal energies. This phenomenon is prevalent in various astrophysical environments such as solar flares (\citealt{Dere1996,Chitta2020,Li2024}), pulsar wind nebulae (\citealt{Meyer2010,Tavani2011}), and gamma-ray bursts (\citealt{Gehrels2009}). \cite{Sweet1958} and \cite{Parker1957} developed a theoretical model, known as the Sweet-Parker model, to describe reconnection processes. In this model, the reconnection rate is determined by $V_{\rm R} \sim V_{\rm A}(\Delta/L) \sim V_{\rm A}S^{-1/2}$, where $S = LV_{\rm A}/\eta$ represents the Lundquist number, $V_{\rm A}$ the Alfv\'en velocity, $L$ and $\Delta$ the length and thickness of the current sheet, as well as $\eta$ the Ohmic resistivity. Given the huge Lundquist number in the astrophysical environment, the Sweet-Parker predicts a low reconnection rate, which contradicts the observations (see \citealt{Dere1996} for an example of fast-flare phenomena). To enhance the reconnection rate, \cite{Petschek1964} proposed an X-type reconnection configuration (known as an X-point model), which converges magnetic field lines to the reconnection region at a sharp angle, resulting in $L$ being so small that it is comparable to $\Delta$. Currently, some observations (\citealt{Shibata1995,Eriksson2004,Drake2006}) and simulations (\citealt{Lin2000, Wang2023}) support this model. Meanwhile, a 2D Petschek-type standard flare model can also explain various macroscale flare characteristics of the Sun (\citealt{Carmichael1964, Sturrock1966, Hirayama1974, Kopp1976}). Although this structure can effectively enhance the reconnection rate, the main challenge is that this structure cannot be maintained on a large astrophysical scale, rapidly collapsing to the Sweet-Parker configuration, as found in MHD simulations (\citealt{Biskamp1996}). 

Given that the magnetic reconnection process is highly dynamic, \citeauthor{Lazarian1999} (\citeyear{Lazarian1999}, hereafter LV99) introduced turbulence to magnetic reconnection, inducing magnetic field wandering to thicken the reconnection layer $\Delta$. Since the wandering of the large-scale magnetic field directly determines the value of $\Delta$, the LV99 theory found that the reconnection rate significantly depends on the turbulence level rather than the plasma resistivity. In the framework of turbulent reconnection model, the fast reconnection rate is given by $V_{\rm R} \approx V_{\rm A}{\rm min}[(l/L)^{1/2},(L/l)^{1/2}]M_{\rm A}^2$, where $l$ is the scale of turbulent eddies. Naturally, for a trans-Alfv\'enic turbulence ($M_{\rm A} \sim 1$), the reconnection rate will reach the Alfv\'en velocity $V_{\rm A}$ on the system size. The 2D numerical testing of the LV99 model demonstrated that the reconnection rate still depends on the resistivity (\citealt{Loureiro2009,Kulpa-Dybel2010}). Note that the LV99 model itself should be 3D in nature due to the 3D dynamic interactions of magnetic fields. By injecting turbulent flow into an anti-parallel magnetic field using the external driving method, \cite{Kowal2009, Kowal2012} performed the first 3D reconnection simulations to test the LV99 model and demonstrated that reconnection is indeed rapid in the presence of turbulence, as it is independent of resistivity. The reason for the difference between 2D and 3D numerical testing is that the 2D and 3D magnetized turbulence numerically exhibit distinct characteristics \cite{Eyink2011}. From the perspective of analytical analysis, \cite{Vishniac2012} predicted that turbulence within the reconnection region is similar to turbulence in a homogeneous system.

Turbulence can also be generated by the inhomogeneous current sheet caused by the presence of magnetohydrodynamic (MHD) instabilities, rendering the imposed external turbulence unnecessary, which is called self-driven (SD) reconnection (see LV99 and \citealt{Lazarian2009b} for details). In the SD reconnection scenario, the inhomogeneity of the current sheet will make the current sheet thicker and the outflow stronger, leading to an increase in turbulence. The reconnection process facilitates turbulence interactions and increases the reconnection rate predicted in LV99. Subsequently, numerical simulations (\citealt{Beresnyak2017, Kowal2017}) verified this theoretical prediction. In the case of incompressibility, \cite{Beresnyak2017} numerically simulated the 3D case of SD reconnection, with the findings as follows: (1) unlike the 2D reconnection, the position of the current sheet changes with the evolution of the system in 3D case; (2) the flux ropes observed in the 3D case are different from the 2D magnetic island exhibiting turbulence within its interior; (3) the number of flux ropes is independent of the Lundquist number, in contrast to \cite{Uzdensky2010}. Moreover, \cite{Beresnyak2017} found that the power spectrum presents the power-law relationship of $E(k) \propto k^{-5/3}$, consistent with the Goldreich-Sridhar (\citeyear{Goldreich1995}, hereafter GS95) theory. 

In the case of the isothermal equation of state (EOS), \cite{Oishi2015} claimed a weak dependence of the width of the current sheet $\Delta$ on the resistivity $\eta$. However, they did not examine the properties of turbulence. \cite{Kowal2017} reported that the reconnection turbulence is similar to the pure MHD turbulence but significantly affected by the flow dynamics induced by reconnection, especially in the low $\beta$ region. Furthermore, \cite{Kowal2020} found that the tearing instability dominates the reconnection rate at the early stage. In contrast, when sufficient turbulence amplitude is reached near the current sheet, the Kelvin-Helmholtz instability dominates turbulence generation over the tearing instability. In the case of the adiabatic EOS, \cite{Huang2016} found a nearly scale-independent anisotropy of the velocity fluctuations, with a slope in the range of $-2.5$ to $-2.3$, in contrast to the theoretical predictions of GS95. In the framework of SD reconnection, our recent work studied the dynamics of magnetic reconnection and the properties of acceleration of the test particle \citep{Liang2023}. We found that the power spectrum of the magnetic field exhibits a steeper slope than the Kolmogorov spectrum, indicating that more magnetic energy accumulates on large scales.

On the one hand, there is a significant contradiction between different numerical simulations of self-driven reconnection. On the other hand, compared to the externally driven reconnection model, the complete theory of self-driven reconnection does not yet exist. Our current work numerically studies the nature of reconnection-driven turbulence. We want to know how different key physical parameters such as the guide field, resistivity and fluid state affect the properties of reconnection turbulence. The structure of this paper is organized as follows. Section \ref{sec:numer} includes the numerical methods and initial setup. Our numerical result is provided in Section \ref{sec:resul}, followed by discussion and summary in Sections \ref{sec:Discu} and \ref{sec:Summa}, respectively.

\section{Simulation setup} \label{sec:numer}
To describe the turbulent magnetic reconnection process, we consider the resistive MHD equations as follows:

\begin{equation} \label{MHD-eq1}
    \frac{\partial \rho}{\partial t} + \nabla \cdot \left(\rho \bm v_{\rm g}\right)=0,
\end{equation}
\begin{equation} \label{MHD-eq2}
    \frac{\partial \bm m}{\partial t}+ \nabla \cdot \left[ \bm{mv}_{\rm g}-\bm{BB}+I\left( p+\frac{\bm B^2}{2}\right)\right]= 0,
\end{equation}
\begin{equation} \label{MHD-eq3}
    \frac{\partial E_{\rm t}}{\partial t} + \nabla \cdot \left[\left(\frac{\rho \bm v_{\rm g}^2}{2}+\rho e +p\right)\bm v_{\rm g} + \bm E \times \bm B\right] = 0,
\end{equation}
\begin{equation} \label{MHD-eq4}
    \frac{\partial \bm B}{\partial t} + \nabla \times \bm E = 0,
\end{equation}
\begin{equation}\label{MHD-eq5}
    \nabla \cdot \bm B = 0,
\end{equation}
representing the continuity, momentum, energy, induction, and solenoidal condition, respectively. Here, $\rho$ denotes the mass density, $\bm v_{\rm g}$ the gas velocity, $\bm m = \rho \bm v_{\rm g}$ the momentum density, $I$ the unit tensor, $e$ the specific internal energy, $p$ the gas pressure, $\bm B$ the magnetic field, and $E_{\rm t} = \rho e +\bm m^2/2\rho+\bm B^2/2$ the total energy density. The evolution of the magnetic field is governed by Faraday's law $\bm E = -\bm v_{\rm g}\times \bm B+\eta \bm J$, where $\bm J = \nabla \times \bm B$ represents the current density and $\eta$ the resistivity coefficient. 

We set the configuration of the initial magnetic field by a Harris-type (\citealt{Harris1962}),

\begin{equation}
    \bm B=B_0 \tanh \frac{y}{w}\bm e_x,
\end{equation}
where $w$ is the initial width of the current sheet and $B_0 = 1.0$ is the magnetic field strength. Magnetic fields are antiparallel along the $X$ direction, resulting in a discontinuity plane placed at $Y=0$. In the case of the adiabatic EOS, we set an initial equilibrium condition (\citealt{Mignone2018,Puzzoni2021})

\begin{equation}
    p = \frac{\beta+1}{2}B_0^2-\frac{\bm B^2}{2}
\end{equation}
to constrain the gas pressure, where the plasma parameter $\beta$ is set to 1.0 (\citealt{Mignone2018}). As a result, the total pressure remains constant throughout the current sheet. For the isothermal cases, we consider the gas pressure of $p=\rho c_{\rm s}^2$ with a sound speed of $c_{\rm s}=1.0$.

We employ the AMUN code (\citealt{Kowal2009, Kowal2012}) to solve Eqs. (\ref{MHD-eq1})--(\ref{MHD-eq5}), where Eq. \eqref{MHD-eq3} will be removed in the case of isothermal states. We perform 3D simulations in physical dimensions of $1.0 \times 1.0 \times 1.0$. Taking into account the numerical resolution of $512^3$ ($1024^3$), we have a grid size of $\delta L \sim 0.002$ ($\delta L \sim 0.001$). In our simulations, we consider periodic boundaries in the $X$ and $Z$ directions and reflective conditions in the $Y$ direction, with the HLLD Riemann solver with WENO reconstruction and third-order RK time stepping. With some fixed parameters such as the uniform background plasma density of $\rho = 1.0$, Alfv\'en velocity $V_{\rm A}=1.0$, and the half-width of the initial current sheet $w \sim 0.01$ (corresponding to about 10 cells), we adjust the variable parameters listed in Table \ref{tab:parameters}. To accelerate simulations, we set an initial velocity disturbance with an amplitude of $V_{\rm eps}$ near the initial discontinuity plane ($Y=0$). When reaching a statistically steady state for the spectral distributions of both kinetic and magnetic energies, we terminate the simulation at the integration time of $t_{\rm final} = 20t_{\rm A}$. 

\begin{table}
\caption{Variable parameters used in our simulation.} 
    \label{tab:parameters}
    \centering
    \setlength{\tabcolsep}{4pt}
    \begin{tabular}{lccccccc}
    \toprule
     Models & $B_{\rm g}$ & $\eta[10^{-4}]$ & $S[10^4]$ & ${V_{\rm eps}}$ & EOS & Resolution\\
     \midrule
     SD1  & 0.0 &1.0  & 1.0 & $0.01V_{\rm A}$  &  ISO & $1024^3$ \\
     SD2  & 0.0 &1.0  & 1.0 & $0.01V_{\rm A}$  &  ISO & $512^3$\\
     SD3  & 0.0 &1.0  & 1.0 & $0.1V_{\rm A}$   &  ISO & $512^3$ \\
     SD4  & 0.0 & 0.164 & 6.1 & $0.1V_{\rm A}$   &  ISO & $512^3$ \\
     SD5  & 0.1 & 0.164 & 6.1 & $0.1V_{\rm A}$   &  ISO & $512^3$ \\
     SD6  & 0.0 & 0.164 & 6.1 & $0.1V_{\rm A}$   &  ADI & $512^3$ \\
     \midrule
    \end{tabular}
    \tablefoot{The physical meanings of symbols are as follows: $B_{\rm g}$--guide field; $\eta$--resistivity coefficient; $S$--Lundquist number; and ${V_{\rm eps}}$--initial velocity  perturbation. The abbreviation ISO and ADI represent isothermal and adiabatic states, respectively. Models SD1 to SD3 correspond to the case of explicit resistivity, and models SD4 to SD6 correspond to the case of implicit resistivity from numerical dissipation.
    }
\end{table}

\section{Simulation results} \label{sec:resul}
\subsection{Dynamic process of turbulent reconnection} \label{subsec:result1}
\begin{figure*}
    \centering
    \includegraphics[width=0.4\linewidth]{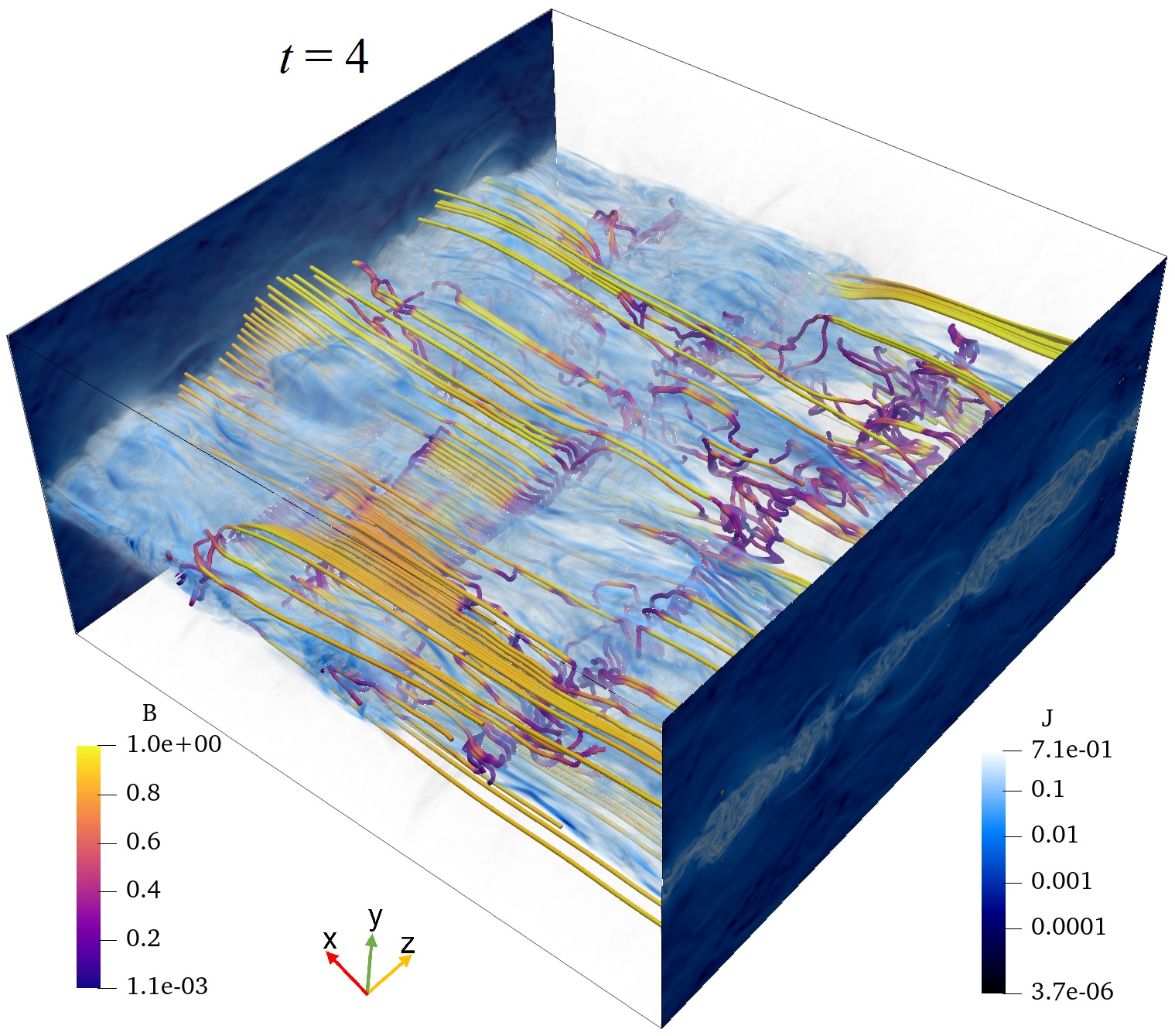}
    \includegraphics[width=0.42\linewidth]{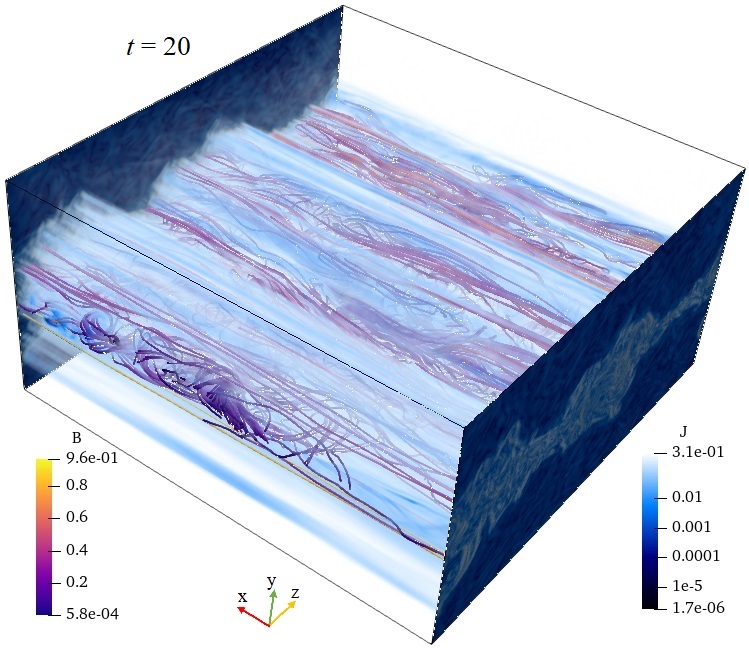}
    \caption{3D view of turbulent reconnection structure at $t=4t_{\rm A}$ and $20t_{\rm A}$. The solid lines represent the magnetic field lines colored with the field strength, filled in the 3D current isosurface with $|J|\ge 0.01$. This 3D view is based on SD2 listed in Table \ref{tab:parameters}.}
    \label{fig:3D-view}
\end{figure*}

Figure \ref{fig:3D-view} visualizes the 3D topological structures of the current sheet at two selected snapshots through transparent volume rendering of the current density isosurface ($|\bm J| > 0.01$), superimposed with magnetic field lines colored by the field strength. From this figure, we see that the current sheet deforms at $t=4t_{\rm A}$ and becomes thicker at $t=20t_{\rm A}$. The dissipated regular magnetic fields ($\langle{B} \rangle \simeq$ 0.9 and 0.7 for $t=4t_{\rm A}$ and $20t_{\rm A}$, respectively) are to increase the fluctuation magnetic fields and kinetic energies in the process of reconnection development. As expected, we also see two distinct magnetic field configurations: (1) highly twisted helical structures in the flux ropes with chaotic magnetic field lines within the reconnection region, which is characteristic of turbulence-dominated magnetic topology configuration; (2) the ordered magnetic field along the background field direction outside the reconnection region has small fluctuations, which is the laminar-dominated configuration. This distinct spatial separation between the turbulent and laminar domains represents a fundamental topological signature of 3D turbulent reconnection. It is because the flux rope merging process drives the turbulence cascades, generating secondary current sheets and effectively limiting the turbulent energy within the current layer (see also \citealt{Dong2022}).

\begin{figure}
    \centering
    \includegraphics[width=0.95\linewidth]{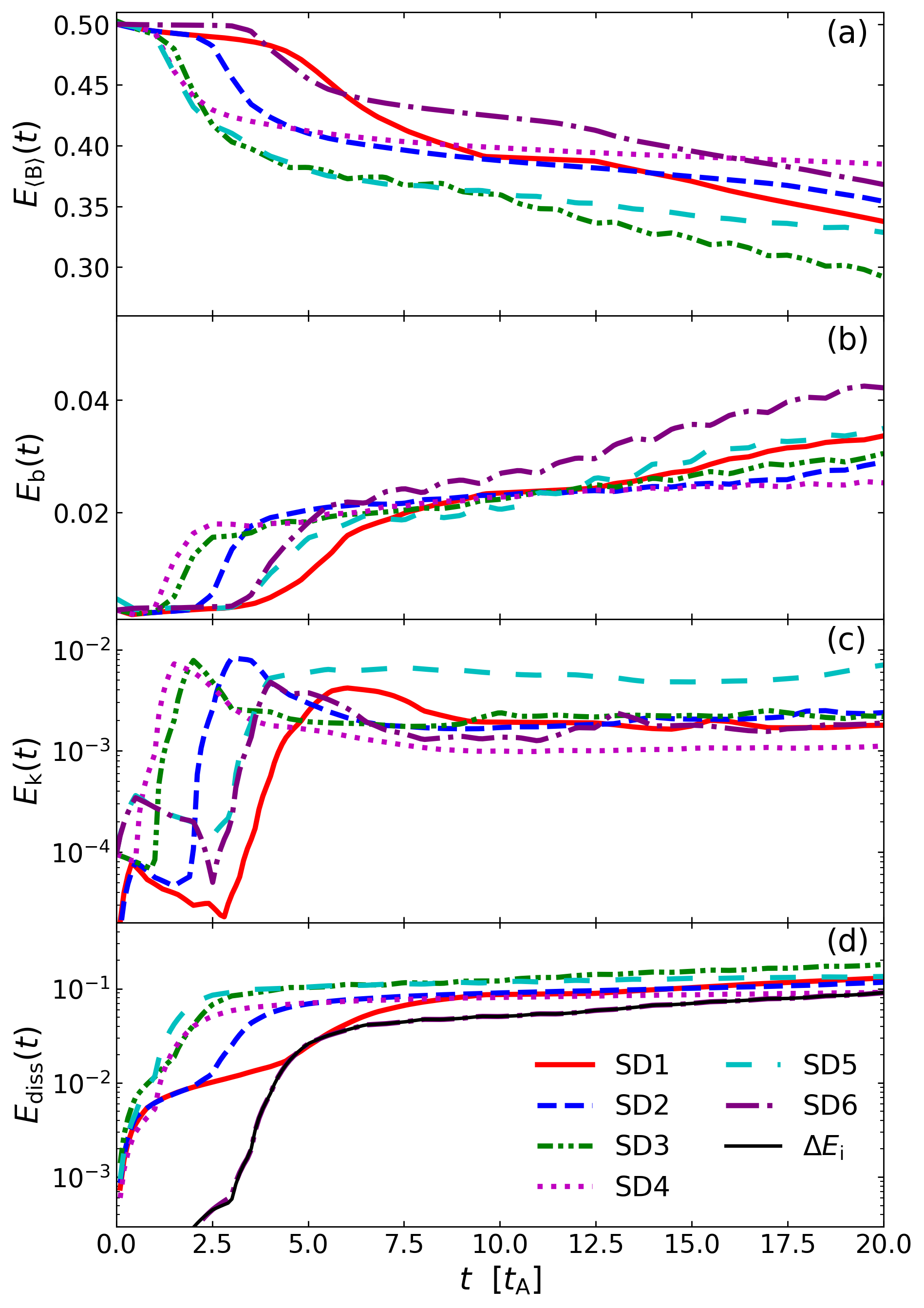}
    \caption{The mean (panel (a)) and fluctuation (panel (b)) magnetic energies, kinetic energy (panel (c)), and dissipated energy (panel (d)) as a function of the evolution time. The black solid line in panel (d) represents the internal energy growth of SD6.}
    \label{fig:time-change-energy}
\end{figure}

To understand exchange process of energy in turbulent magnetic reconnection, Figure \ref{fig:time-change-energy} plots temporal variation of the magnetic (Figs. \ref{fig:time-change-energy}a and b), kinetic (Fig. \ref{fig:time-change-energy}c), and dissipated (Fig. \ref{fig:time-change-energy}d) energies for all models, where the magnetic energy includes the mean $E_{\rm {\langle B\rangle}}$ (Fig. \ref{fig:time-change-energy}a) and fluctuation $E_{\rm b}$ (Fig. \ref{fig:time-change-energy}b) components, and dissipated energy $E_{\rm diss}(t) = E_{\rm {\langle B\rangle}}(t=0) + E_{\rm k}(t=0) - E_{\rm {\langle B \rangle}}(t)-E_{\rm b}(t) - E_{\rm k}(t)$, where the internal energy is defined as $E_{\rm i} = \frac{p}{\Gamma-1}$ ($\Gamma$ is the adiabatic index), and its growth is written as $\Delta E_{\rm i}(t) = E_{\rm i}(t)-E_{\rm i}(t=0)$. As we can see, all the cases exhibit a similar evolutionary trend, with only differences in temporal progression. The overall evolution process can be divided into three phases: (1) the initial phase of SD reconnection, resulting from instability triggered by the initial velocity perturbation within the current sheet; (2) the fast dissipation/excitation phase, the regular magnetic energy released by reconnection being transferred into kinetic energy and fluctuation magnetic energy, with enhancing the turbulence inside the current sheet to promote the magnetic reconnection process; (3) the subsequent gradual relaxation/excitation phase, showing an approximately dynamic equilibrium between the dissipation of regular magnetic energies and the increasing of fluctuation kinetic and magnetic energies. 

Taking SD1 as an example, we further clarify three phases mentioned above: (1) for the first phase before $t \sim 3t_{\rm A}$, the regular magnetic energy $E_{\rm {\langle B\rangle}}$ slowly dissipates and the fluctuation magnetic energy $E_{\rm b}$ nearly remains unchanged (Figs. \ref{fig:time-change-energy}a and b), while kinetic energy $E_{\rm k}$ also decreases due to numerical dissipation (Figs. \ref{fig:time-change-energy}c and d), the dissipated energy $E_{\rm diss}$ significantly increases. At this stage, initial velocity perturbation triggers instabilities, leading to the beginning of reconnection and the dissipation of magnetic energy within the current sheet; (2) for the second phase during $3t_{\rm A} \le t \le 7t_{\rm A}$, the energy of the regular magnetic field $E_{\rm {\langle B\rangle}}$ significantly decreases, the fluctuation magnetic field $E_{\rm b}$ and kinetic field $E_{\rm k}$ increases rapidly, because of the emergence of turbulence to enhance the reconnection level; (3) for the third phase after $t \sim 7t_{\rm A}$, $E_{\rm {\langle B\rangle}}$ and $E_{\rm b}$ gradually decrease and increase, respectively, while $E_{\rm k}$ and $E_{\rm diss}$ remain an approximate plateau, with the ratio of kinetic energy to dissipated one $E_{\rm k}/E_{\rm diss}  \sim 2\%$ at $t = 20t_{\rm A}$, showing an approximate relationship of $\Delta E_{\rm {\langle B\rangle}} \simeq \Delta E_{\rm k}+ \Delta E_{\rm b}+\Delta E_{\rm diss}$.

In addition, we note that the internal energy evolution ($\Delta E_{\rm i}(t)$ vs. $t$) of the adiabatic model SD6 almost coincides with that of the dissipated energy (see SD6 in Fig. \ref{fig:time-change-energy}d), indicating that the dissipated energy is mainly converted into the thermal energy of the gas. This result can help to understand observations in solar flares by \cite{Kontar2017}, who pointed out that the magnetic energy released by reconnection is dissipated into gas thermal energy via turbulent interactions, and turbulence kinetic energy accounts for about $1\%$ of the magnetic energy released by reconnection (i.e., dissipated energy).

\subsection{Direction-dependent spectral distribution}
\begin{figure*}
    \centering
    \includegraphics[width=0.9\textwidth,bb=10 10 800 350]{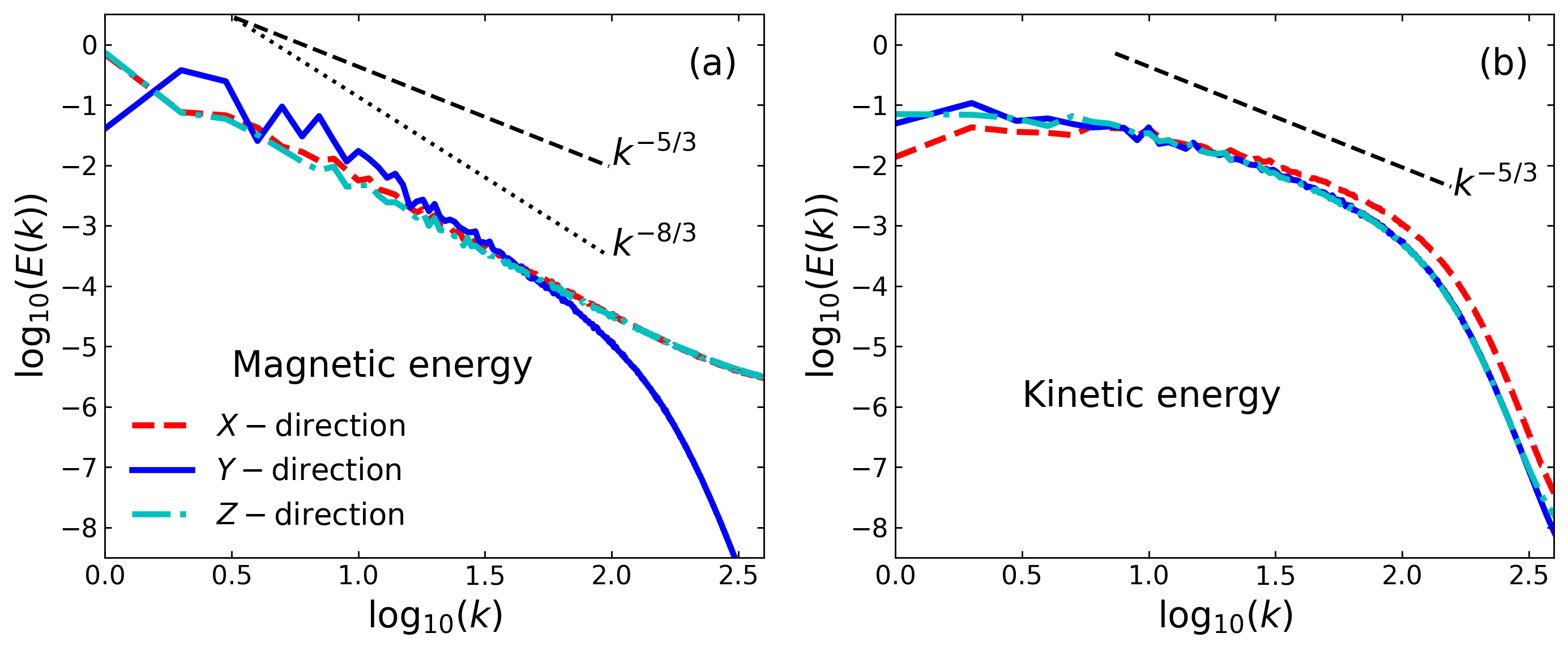}
    \includegraphics[width=0.9\textwidth,bb=10 0 890 350]{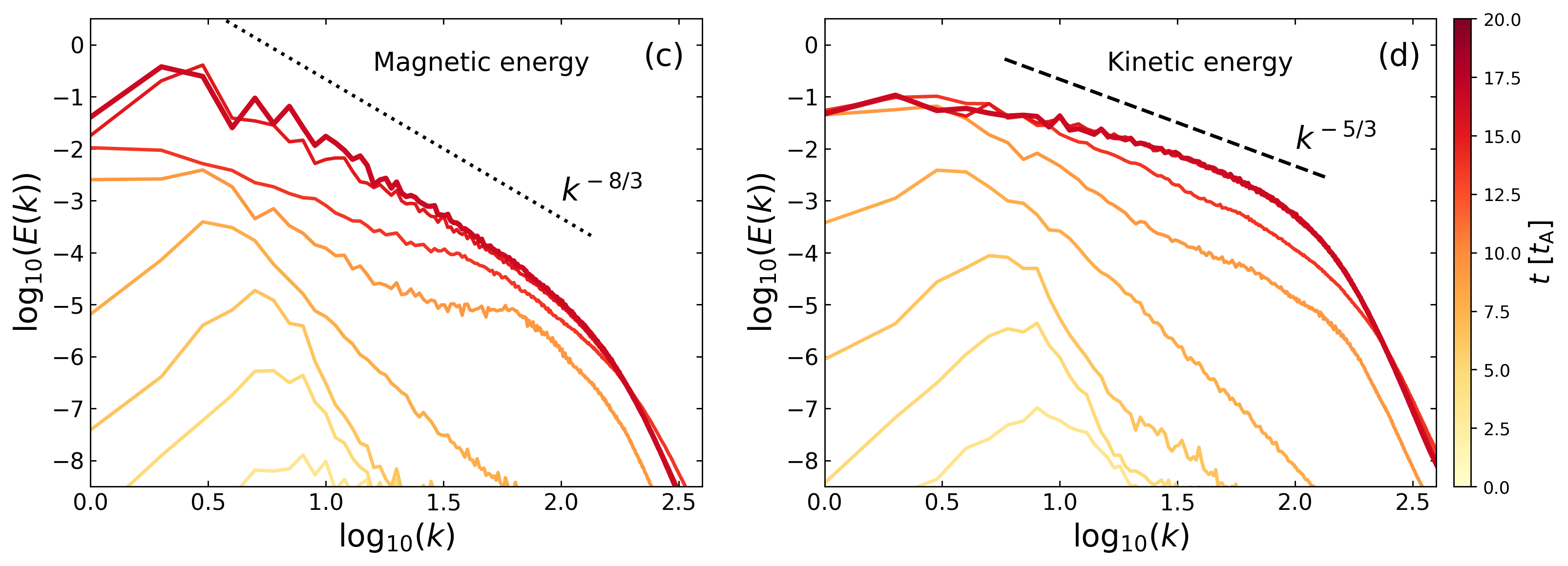}
    \caption{Spectral distributions of magnetic (left column) and velocity (right column) fields. Upper row: spectral distributions in the three axis directions at $t=20t_{\rm A}$. Lower row: spectral evolutions in the process of turbulence development, where the color bar shows the simulation time in units of Alfv\'en time $t_{\rm A}$. These results are based on high resolution simulations of SD1.}
    \label{fig:PS-direction-time}
\end{figure*}

To investigate the inhomogeneity of reconnection turbulence, we adopt the methodology proposed by \cite{Beresnyak2017} for spectral calculations defined as

\begin{equation} \label{PS}
  E(k_l) = L^{-1}\int \hat f(\bm k_l) \hat f^*(\bm k_l)~d\phi ~dl
\end{equation}
where $l$ and $k_l$ represent the integration direction ($X$, $Y$ and $Z$) and the wave vector in the plane perpendicular to that direction, respectively, and $\hat f(\bm k_l)$ denotes the Fourier transform of $\bm v$ or $\bm B$. Here, we have $k_l=(k_y,k_z)$, $(k_x,k_z)$ and $(k_x,k_y)$ for the $X$, $Y$ and $Z$ directions, respectively. Based on Eq. (\ref{PS}), we first analyze the spectral distribution characteristics at the end time ($t \sim 20t_{\rm A}$) along three coordinate axis directions. 

As shown in Fig. \ref{fig:PS-direction-time}, we present the spectral distributions of magnetic (panel (a)) and kinetic (panel (b)) energies in the $X$, $Y$, and $Z$ directions. For the spectral distributions of magnetic energy, we can see from Fig. \ref{fig:PS-direction-time}a that spectral distributions in the $Y$ direction show a steep slope of $E \propto k^{-8/3}$, which indicates the power more concentrated in the large-scale (small wavenumber) range. In this direction, that is, the $X-Z$ plane parallel to the current sheet, the turbulent magnetic field (small power) within the reconnection region can be well separated from the undisturbed magnetic field (large power) outside the reconnection region. Spectral distributions in the $X$ and $Z$ directions, the $Y-Z$ and $X-Y$ planes perpendicular to the current sheet, show a shallow slope between $E\propto k^{-8/3}$ and $E\propto k^{-5/3}$ that extends to a large wavenumber (without significant dissipation process of magnetic energies at small scales). Since both the turbulent magnetic field (within and surrounding reconnection region) and the undisturbed magnetic field (away from reconnection region) are included in the same layer when calculating the power spectrum, the small-scale fluctuations of reconnection turbulence within the reconnection region are masked by the surrounding strong, regular/large-scale magnetic fields. Therefore, the power spectra in the $X$ and $Z$ directions do not fully reveal the cascading nature of the reconnection turbulence. The asymmetric phenomenon of spectral distributions reveals well the spatial inhomogeneity of turbulent magnetic field distributions in the reconnection turbulence (see also \citealt{Zhang&Liang2024} for synthetic synchrotron observations). For the steep spectral feature of the magnetic field, we speculate that the inhomogeneity of the magnetic field and plasma wave instabilities may lead to an anomalous resistivity (e.g., \citealt{Numata2002,Silin2005,Wu2010}), which may indirectly result in the steepening of the power spectrum by affecting the energy cascade processes.

Spectral distributions of the kinetic energy (see Fig. \ref{fig:PS-direction-time}b) exhibit different characteristics from those of the magnetic field. They follow the classical Kolmogorov spectrum of $E\propto k^{-5/3}$, except that the spectrum in the $X$ direction becomes slightly shallower in the large-scale range, which may be due to the effects of the outflow. It shows that the inhomogeneity of the turbulent flow driven by the reconnection is mainly reflected in the turbulent magnetic field, while the turbulent velocity remains uniform. To effectively reveal the nature of reconnection turbulence, we will discuss only the power spectral properties in the $Y$ direction below.

To trace the evolution process of reconnection turbulence, we plot the time-dependent evolution of spectral distributions of magnetic and kinetic energies in Figs. \ref{fig:PS-direction-time}c and d, respectively. With the increase of evolutionary time, the amplitudes of the magnetic and velocity spectra gradually increase to a quasi-saturated state. In the early stages of evolution, the spectra of velocity and magnetic field only distribute in the large-scale range (about $k < 10$). Subsequently, the distribution of the spectrum extends to a wider range of wavenumbers. In the late stages, they converge to the scaling of $E(k) \propto k^{-8/3}$ for the magnetic field and $E(k) \propto k^{-5/3}$ for velocity.  We note that the spectra $t \sim 20 t_{\rm A}$ (the top red lines) are almost the same as those at $t \sim 15t_{\rm A}$ (the penultimate red lines), indicating that the turbulence has fully developed. Therefore, our subsequent analysis of the turbulence characteristics will focus on the quasi-steady-state at $t = 20t_{\rm A}$.

\subsection{Influence of variable parameters on reconnection turbulence}
\begin{figure*}
    \centering
    \includegraphics[width=0.9\linewidth]{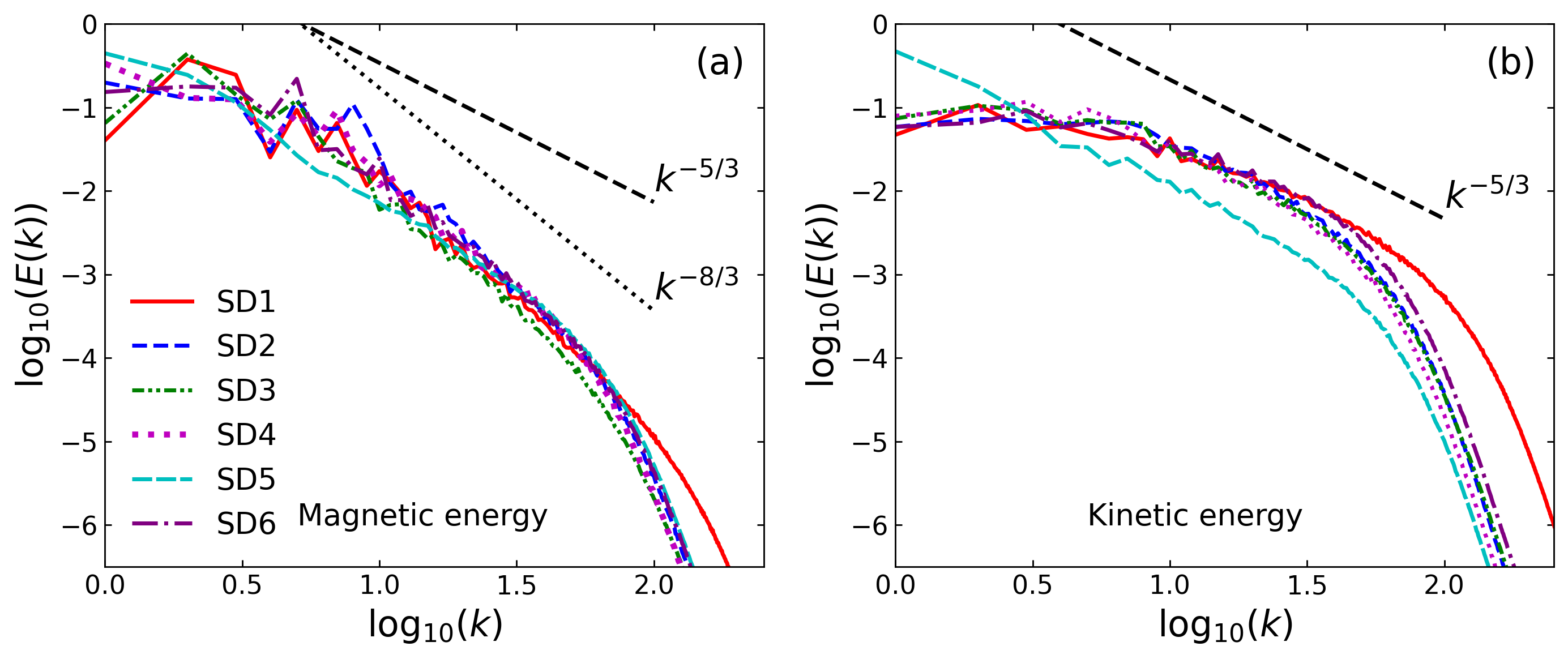}
    \caption{Power spectral distributions of magnetic (panel (a)) and velocity (panel (b)) fields normalized in their mean power at $t=20t_{\rm A}$ for all models SD1 to SD6 listed in Table \ref{tab:parameters} .}
    \label{fig:ps-all}
\end{figure*}

To evaluate the impact of numerical convergence on our numerical results, we ran twice with the same parameter setting (SD1 for $1024^3$ and SD2 for $512^3$) to compare the spectral distributions. Figure \ref{fig:ps-all} shows spectral distributions of the magnetic (panel (a)) and kinetic (panel (b)) energies for all cases at $t=20t_{\rm A}$. As we can see, spectral distributions of magnetic and kinetic energies display consistent statistical behavior in the inertial range, with an extended tail at large wavenumber for SD1. In other word, the simulation for the SD1 exhibits a marginally extended inertial range, as expected that high-resolution simulations can resolve small-scale structures of reconnection turbulence. It confirms that the resolution of $512^3$ can capture the spectral feature of reconnection turbulence cascade. Therefore, to save computing resources, we will perform studies of parameter changes in $512^3$ resolutions below.

Next, we explore the effect of initial velocity perturbations and explicit resistivity settings on the reconnection turbulence. Due to the lack of differences in spectral distributions between SD2 and SD3, we suggest that the initial velocity perturbation has no significant effect on the reconnection turbulence. Similarly, since SD3 and SD4 exhibit nearly identical spectral distributions, we confirm that explicit resistivity does not significantly influence turbulent reconnection dynamics, in agreement with the theoretical predictions of LV99 (e.g., see \citealt{Kowal2009,Kowal2017,Beresnyak2017} for similar results). In addition, we study the influence of a non-zero guide field on reconnection turbulence (see SD5 of Fig. \ref{fig:ps-all}a). For SD5, we see that the magnetic spectrum exhibits an approximate slope of $E\propto k^{-8/3}$, with a softening feature in the range of the large wavenumber of $k< 10$, and the velocity spectrum shows a scaling of $E\propto k^{-5/3}$. Consistent with recent 3D MHD simulations (\citealt{Xiong2024}) and magnetotail observations (\citealt{Fu2017,Huang2023}), the existence of the guiding field can promote the transformation of small-scale reconnection points into larger-scale magnetic flux rope structures(\citealt{Fu2017,Huang2023}).

\begin{figure*}
    \centering
    \includegraphics[width=0.45\linewidth]{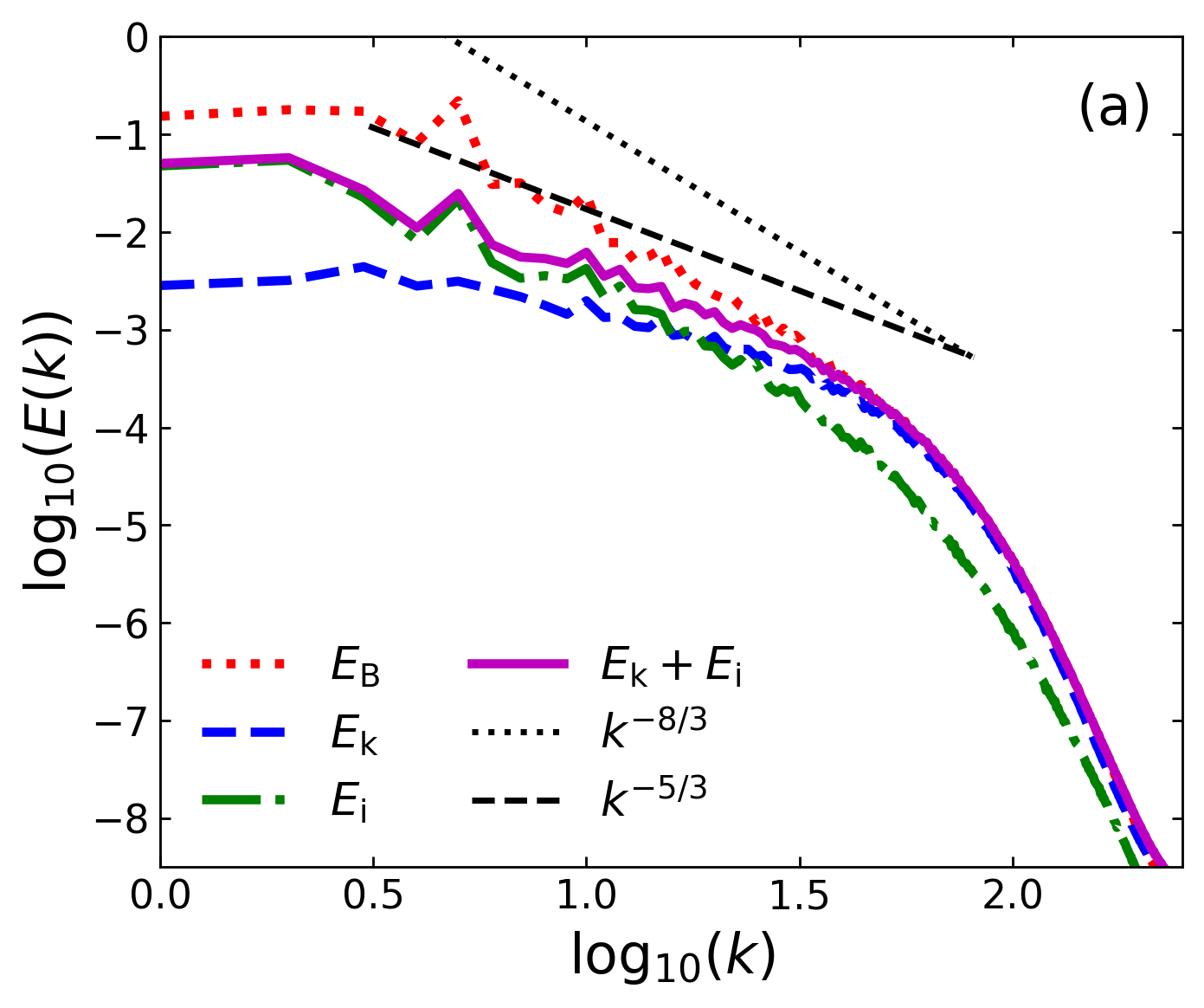}
    \includegraphics[width=0.45\linewidth]{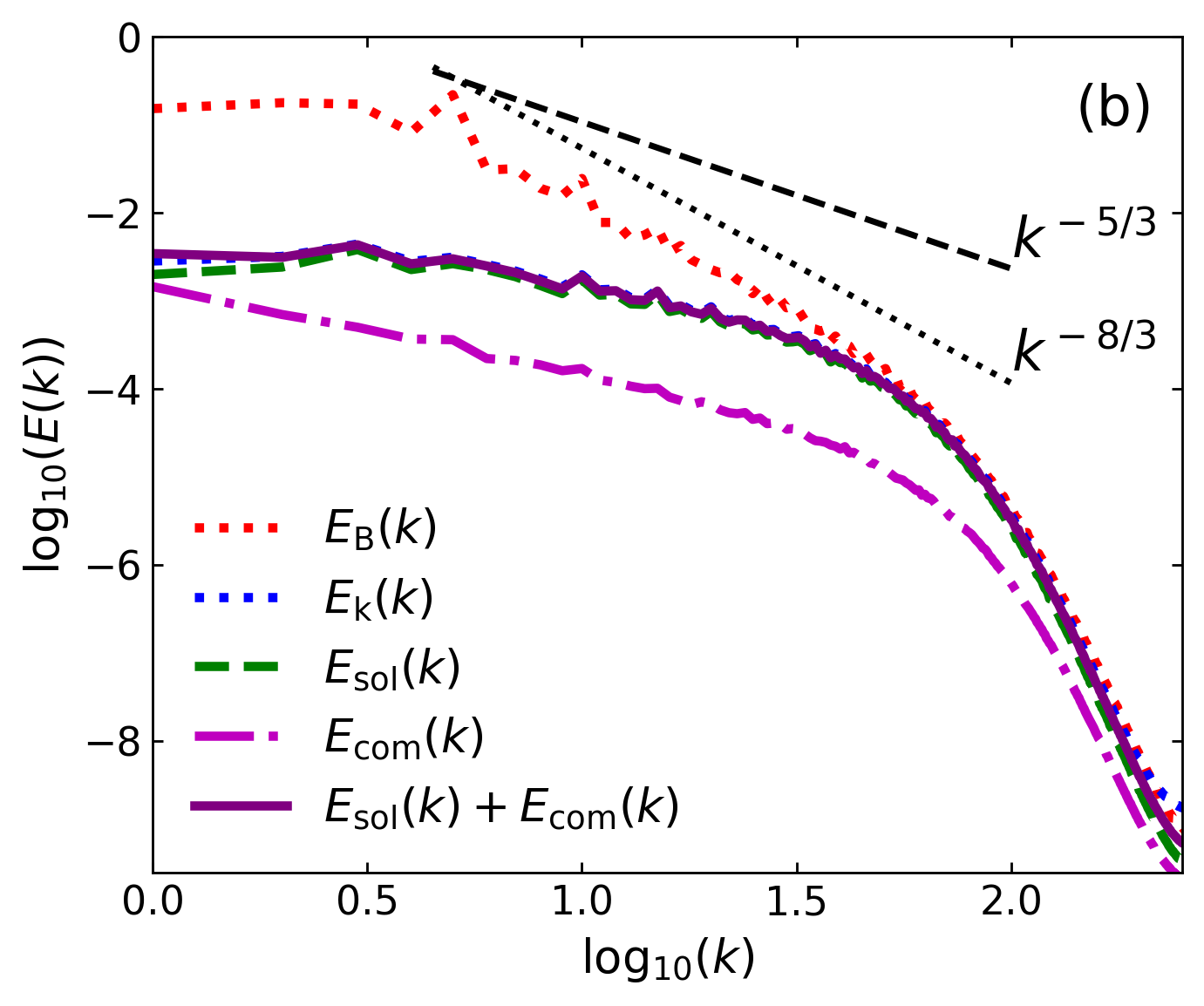}
    \caption{Distributions of magnetic and kinetic energies compare with distributions of internal energies (panel (a)), as well as the solenoidal and compressive components of kinetic energies (panel (b)). Simulations are from the adiabatic state (SD6 listed in Table \ref{tab:parameters}) at $t=20t_{\rm A}$.
    }
    \label{fig:PS-adi-EOS}
\end{figure*}

As shown in Fig. \ref{fig:PS-adi-EOS}a, we also investigate the influence of the EOS on turbulent reconnection by analyzing the spectral distributions of the magnetic, kinetic, and internal energies. From this figure, we see spectral distributions of magnetic and internal energies of $E\propto k^{-8/3}$, and the kinetic energy spectrum $E\propto k^{-5/3}$. Remarkably similar spectral distributions from magnetic and internal energies suggest a close connection between magnetic dissipation and gas heating. In addition, a least-squares fitting yields a parabolic relation of $p = 0.51B^2 - 0.01B + 1$ (not shown), which implies that magnetic annihilation during turbulent reconnection enables efficient energy conversion via Joule heating (LV99).

\subsection{Compressibility of reconnection turbulence} \label{Helmholtz-dec}
To elucidate the compressibility of reconnection turbulence, we perform a Helmholtz decomposition of the velocity field ($\bm v = \bm v_{\rm sol} + \bm v_{\rm com}$) and compute the spectral distributions for two components (\citealt{Zhdankin2017}), which are defined as $E_{\rm sol}(k)=|\bm k \times \bm {\hat v}(k)|^2$ and $E_{\rm com}(k)=|\bm k \cdot \bm {\hat v}(k)|^2$. Figure \ref{fig:PS-adi-EOS}b shows the spectral distributions of the magnetic field, velocity field, solenoidal, and compressive components. As we can see, although the solenoidal $E_{\rm sol}(k)$ and compressive $E_{\rm com}(k)$ components all follow $E\propto k^{-5/3}$, the amplitude of the compressive component is much smaller than that of the solenoidal component, with the compressive component $E_{\rm com}(k)$ contributing only $\sim 6.5\%$ to the kinetic energy. Therefore, for the parameter settings of SD6, we suggest that the reconnection turbulence is in a quasi-incompressible state, for which non-reconnection MHD simulations give the proportion of compressible components being less than $10\%$ (\citealt{Cho2003}).

\subsection{Anisotropy of reconnection turbulence} 
\begin{figure*}
    \centering
    \includegraphics[width=0.48\linewidth]{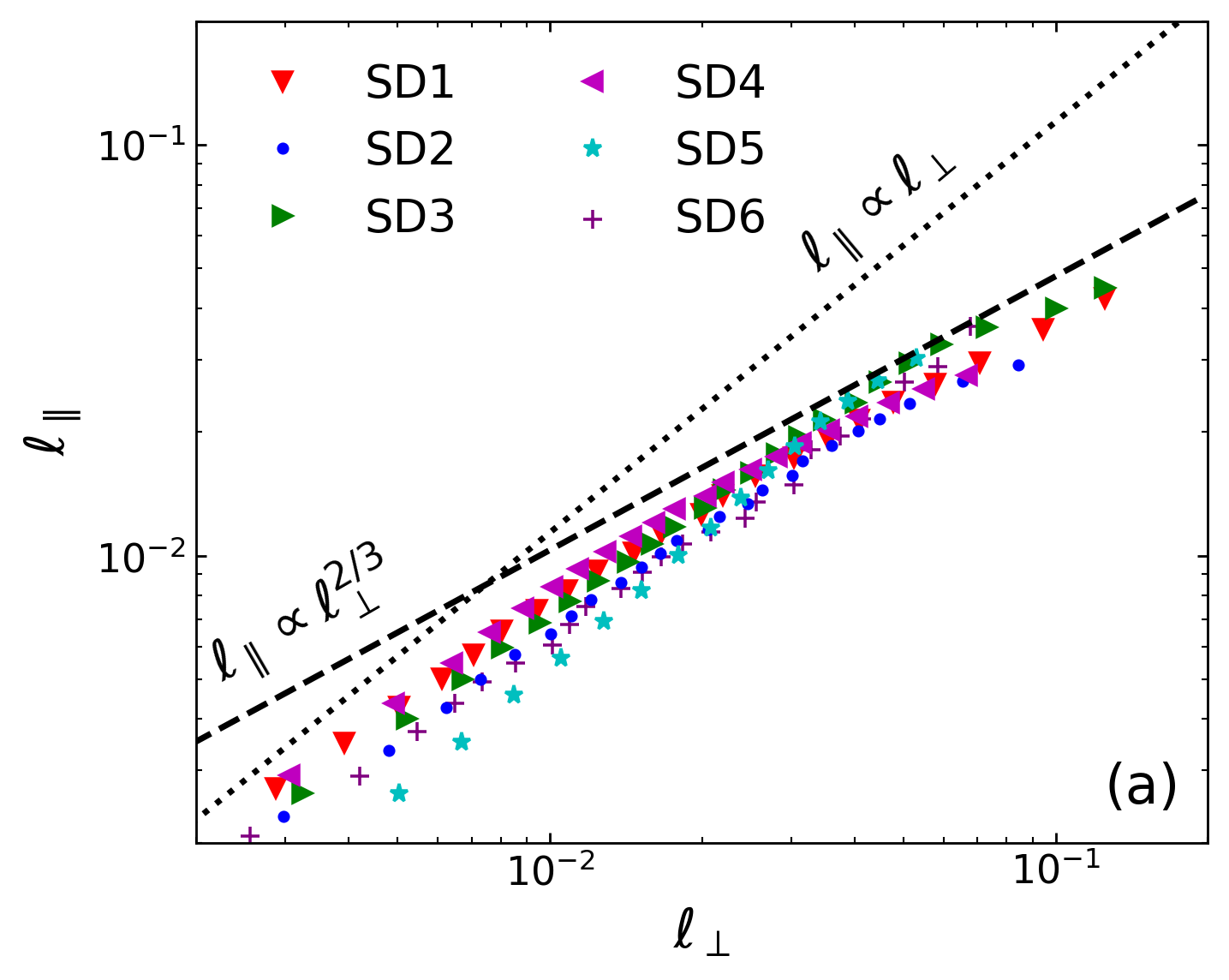}
    \includegraphics[width=0.47\linewidth]{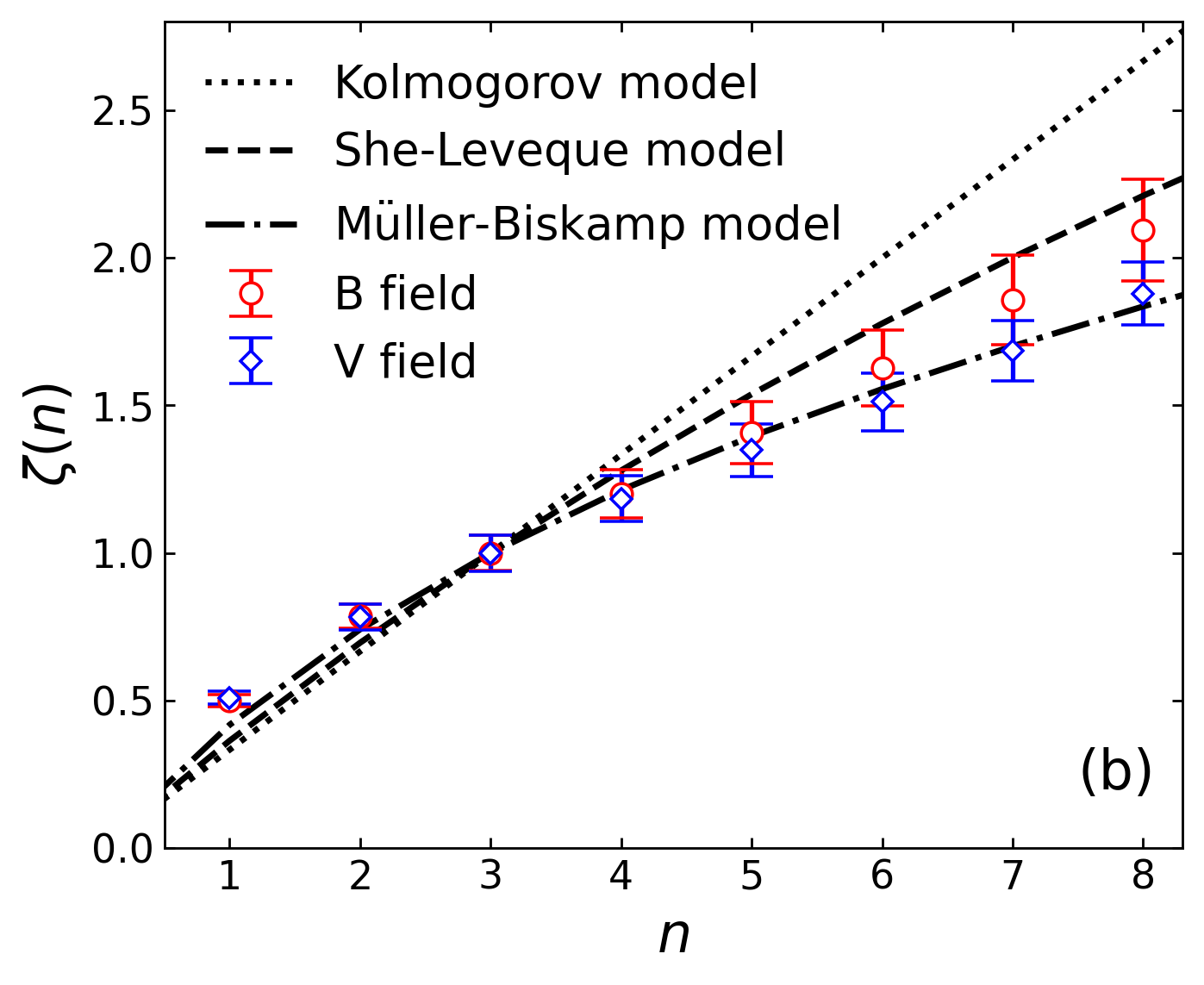}
    \caption{Scale-dependent anisotropy of velocities for all models (panel (a)) and intermittency of magnetic and velocity fields (panel (b)). Panel (a): The coordinates $\ell_{\perp}$ and $\ell_{\parallel}$ represent the perpendicular and parallel scales of the eddies along the local magnetic field, respectively. Panel (b): The scaling exponent $\zeta$ vs. the order of structure functions. Our numerical results with error bars estimated from the standard deviation compare with theoretical predictions of intermittency models: Kolmogorov, She-Leveque, and M\"uller-Biskamp. Numerical results that we analyzed $t=20t_{\rm A}$ are based on SD1 listed in Table \ref{tab:parameters}.
    }
    \label{fig:SF}
\end{figure*}

Figure \ref{fig:SF}a provides the scale-dependent anisotropy for all models listed in Table \ref{tab:parameters}, with black dotted and dashed lines representing the isotropic and anisotropic scalings, respectively. As is seen, the most of models exhibit the anisotropic scaling of $\ell_{\parallel} \propto \ell_{\perp}^{2/3}$ at large scales, except that the SD5 with a non-zero guide field has an isotropic relationship of $\ell_{\parallel} \propto \ell_{\perp}$ throughout the whole spatial region. It indicates that the presence of a guide field affects the spatial anisotropy of the reconnection turbulence. Compared SD1 to SD2, we found that the anisotropic scaling for SD1 is stronger than that of SD2 at small scales, suggesting that the reconstruction of anisotropic relationships depends on numerical resolution. Since our current simulations have a uniform grid distribution, the geometric constraint of reconnection turbulence poses a challenge to high-precision simulations (see \citealt{Wang2025ApJ} for the high-resolution simulation of the coronal mass ejection). As we can see in Fig. \ref{fig:3D-view}, the reconnection turbulence region occupies only a small portion of the entire space in the direction perpendicular to the current sheet. We expect that simulations of non-uniform meshes may be an effective way to capture the anisotropy of reconnection turbulence, which will be explored in future work.

\subsection{Intermittency of reconnection turbulence}
To investigate the intermittency of reconnection turbulence, we use the extended self-similarity (\citealt{Benzi1993}) to obtain the scaling exponent ($\zeta(n)$) of the $n$th-order structure function that is normalized by the scaling exponent of the third-order structure function. Figure \ref{fig:SF}b presents the intermittency of magnetic and velocity fields for SD1 at $t=20t_{\rm A}$. The dotted, dashed, and dot-dashed lines represent theoretical results from the Kolmogorov (\citeyear{Kolmogorov1941}, $\zeta(n)=\frac{n}{3}$), She-Leveque (\citeyear{She1994PRL}, $\zeta(n)=\frac{n}{9}+2[1-(\frac{2}{3})^{n/3}]$) and M\"uller-Biskamp (\citeyear{Muller2000PRL}, $\zeta(n)=\frac{n}{9}+1-(\frac{1}{3})^{n/3}$) models, respectively. As we can see, the velocity and magnetic fields exhibit distinct scaling behaviors. The velocity field scaling at higher orders ($n>3$) aligns with the M\"uller-Biskamp model corresponding to a 2D sheet-like structure. This finding is consistent with \cite{Wang2025ApJ}, in which examination of the reconnection core morphology in the current sheet reveals a preference for 2D sheet-like structures. The magnetic field scaling exponents fall between the results of She-Leveque and M\"uller-Biskamp models. Compared to the velocity field, it means that the intermittency of the magnetic field is weaker than that of the velocity field.

In the case of lower orders ($n$=1 and 2), both velocity and magnetic field scaling exponents slightly exceed the results from all three models. This deviation may arise because classical turbulence theories consider energy cascades from large to small scales. Here,  turbulent magnetic reconnection locally concentrates energy release (through current sheet disruption or magnetic island coalescence). Such an inhomogeneous energy injection could modify the scaling laws (\citealt{Beresnyak2019}). Furthermore, rapid merging of flux ropes (or magnetic islands in 2D) during reconnection may also enhance the small-scale energy cascade and potentially suppress intermittency (\citealt{Loureiro2012}).

\section{Discussion} \label{sec:Discu}
In this work, we conducted simulations of SD turbulent reconnection to investigate the properties of reconnection turbulence. Our simulations have incorporated the resistivity effect but have neglected possible viscosity. This is because in magnetic reconnection research, magnetically dominated plasmas are generally considered (e.g., the solar corona and magnetopause; see \citealt{Lazarian2020}), where resistive dissipation predominantly governs energy conversion within current sheets. This resistive term in the generalized Ohm's law constitutes the primary mechanism for magnetic diffusion (\citealt{Priest2000}), an assumption supported by extensive observational evidence, including solar flares (\citealt{Su2013}) and in situ magnetospheric measurements (\citealt{Burch2016}). Furthermore, energetic transfer in the inertial range is primarily governed by Alfv\'enic wave interactions (\citealt{Kraichnan1965}), which are insensitive to dissipation mechanisms (\citealt{Muller2005}).

Our numerical results indicate that explicit resistivity has no significant effect on the properties of reconnection turbulence, at least in the quasi-steady-state phase (see Sect. \ref{subsec:result1}), which is consistent with the expectation of LV99 and the results of numerical experiments (e.g., \citealt{Kowal2009, Kowal2012, Kowal2017}). In the early stage of the reconnection evolution, the resistivity may play a role when the instability of the tearing mode dominates the reconnection process (\citealt{Huang2010, Huang2016}). In that case, the influence of different Lundquist numbers should be considered, especially for plasmas above and below the critical Lundquist number ($S_{\rm crit}\sim 10^4$, see \citealt{Samtaney2009}). However, once the system transitions to fully developed turbulent reconnection, the effects of resistivity on the reconnection rate can be ignored (\citealt{Kowal2009, Kowal2020, Lazarian2020}).

In the case of the guiding field, our results show that the anisotropy of magnetic energy is significantly different from that of other cases. This difference highlights the influence of the guide field on the reconnection turbulence. It is incompatible with the results provided by \cite{Kowal2009}, who found that the guide field has no significant impact on the reconnection process. The reason may be that with the high injection rate of turbulence ($P_{\rm inj}\sim 1.0$) in \cite{Kowal2009}, the turbulence cascade is strong enough. An appropriate guide field can provide an angle of inclination for the antiparallel magnetic field lines to facilitate the reconnection rate (see \citealt{Liang2023} for SD  reconnection; and \citealt{Xu2023} for a theoretical explanation), resulting in an effective acceleration of particles (\citealt{Liang2023,Zhang2023}). This difference requires more numerical simulation work to test.

The reconstruction of scale-dependent anisotropy depends on high-resolution numerical simulations. Deviation of anisotropic relationships at small scales (see \citealt{Kowal2017} for similar results) may be caused by numerical dissipation. Therefore, we indeed need higher-resolution simulations to confirm whether the consistency between the anisotropic scaling of reconnection turbulence and the GS95 theory is robust. We expect that adaptive mesh refinement may be an effective method for achieving high-resolution simulations to capture the anisotropy of reconnection turbulence. We will explore this in future work.

\cite{Wang2025ApJ} analyzed the reconnection core within the current sheet, indicating that the reconnection core presents a 2D sheet-like structure, which is consistent with our intermittent result of the velocity field (Fig. \ref{fig:SF}b). A notable finding is that, in reconnection turbulence, the velocity field has a stronger intermittency than the magnetic field; this result contrasts with pure MHD turbulence (e.g., \citealt{Cho2003a, Haugen2004, Yoshimatsu2011, Mallet2016}). This suggests that the cascade processes of the turbulent magnetic field are constrained by reconnection processes within the current sheet. 

Different from the well-known Kolmogorov spectrum of $E\propto k^{-5/3}$, our simulations show an anomalous steep spectrum of the magnetic field of $E\propto k^{-8/3}$ along the $Y$-axis direction, which is a better representation of the spectrum of anisotropic reconnection turbulence. This result is similar to the spectral shape of electron MHD turbulence on kinetic scales in the case of strong magnetic fields (\citealt{Meyrand2013}), arising from the strong suppression of the energy cascade along the magnetic field. In the case of reconnection turbulence, reconnection outflows interacting with the magnetic field may result in a negligible parallel (with regard to the local magnetic field) cascade. Therefore, we claim that the anomalous steep spectrum may not be an isolated case across different spatial scales.

\section{Summary} \label{sec:Summa}
Focusing on the properties of reconnection-driven turbulence in SD turbulent magnetic reconnection, we explored the influence of different factors (such as the guide field, resistivity, numerical resolution, and fluid state) on the dynamics and turbulence of reconnection. The main results are briefly summarized as follows:

\begin{enumerate} 
\item The evolution of turbulent reconnection can be simply divided into three phases: initial, fast dissipation, and relaxation phases. The dissipation of the antiparallel, regular magnetic fields leads to an increase in turbulent magnetic and kinetic energies.

\item Due to the inhomogeneity of reconnection turbulence, the power spectrum of the turbulent magnetic field presents a direction-dependent spectral distribution, and the power spectrum of turbulent velocity is almost direction-independent.

\item Velocity spectrum shows the expected scaling of $E\propto k^{-5/3}$ corresponding to the classic Kolmogorov type, while the slope of the magnetic field spectrum is steeper. 

\item Explicit resistivity and initial velocity disturbance amplitude have no significant effect on the properties of reconnection turbulence, such as spectra and anisotropy, which are consistent with the theoretical prediction of LV99. The presence of a guide field results in an isotropic cascade of reconnection turbulence, which is significantly different from the cascade pattern of pure MHD turbulence.

\item Reconnection turbulence presents a quasi-incompressibility in the adiabatic states, with energy distributions dominated by the solenoidal component of the velocity. The intermittency of the velocity field is stronger than that of the magnetic field.
\end{enumerate}

\begin{acknowledgements}
We thank the anonymous referee for valuable comments that significantly improved the quality of the paper. This work is supported in part by the High-Performance Computing Platform of Xiangtan University. J.F.Z. is grateful for the support from the National Natural Science Foundation of China (No. 12473046), and the Hunan Natural Science Foundation for Distinguished Young Scholars (No. 2023JJ10039). N.Y.Y. is grateful for the support from the National Natural Science Foundation of China (No. 12431014) and the Project of Scientiﬁc Research Fund of the Hunan Provincial Science and Technology Department (2024ZL5017). S.M.L. and N.N.G. are grateful for the support from the Xiangtan University Innovation Foundation for Post-graduate Students (Nos. XDCX2024Y164 and XDCX2025Y257). 
\end{acknowledgements}
\vspace{5mm}

\bibliographystyle{aa}
\bibliography{references}

\end{document}